\documentclass{aa}

\usepackage{graphicx}
\usepackage{txfonts}
\usepackage{braket}
\usepackage{bm}
\usepackage[colorlinks,allcolors=blue]{hyperref}
\usepackage[a]{esvect}
\usepackage{xcolor}
\makeatletter

\renewcommand*\aa@pageof{, page \thepage{} of \pageref*{LastPage}}
\def\mtrx#1{\bar{\bar{{\bf #1}}}}
    
\makeatother
\usepackage{amstext}
\usepackage[normalem]{ulem}

\begin{document}

\title{New Periodograms Separating Orbital Radial Velocities and Spectral Shape Variation}

\author{A. Binnenfeld\inst{\ref{inst1}} \and S. Shahaf\inst{\ref{inst2}} \and Richard I. Anderson\inst{\ref{inst3},\ref{inst4}} 
\and S. Zucker\inst{\ref{inst1}}}

\institute{Porter School of the Environment and Earth Sciences, Raymond and Beverly Sackler Faculty of Exact Sciences, 
Tel Aviv University, Tel Aviv, 6997801, Israel \\
\email{avrahambinn@gmail.com}\label{inst1}
\and
Department of Particle Physics and Astrophysics, Weizmann Institute of Science, Rehovot 7610001, Israel
\label{inst2}
\and
Institute of Physics, Laboratory of Astrophysics, \'Ecole Polytechnique F\'ed\'erale de Lausanne (EPFL), Observatoire de Sauverny, 1290 Versoix, Switzerland
\label{inst3}
\and
European Southern Observatory, Karl-Schwarzschild-Str. 2, 85748 Garching, Germany
\label{inst4}}

% These dates will be filled out by the publisher
\date{Accepted XXX. Received YYY}

\abstract{
We present new periodograms that are effective in distinguishing Doppler shift from spectral shape variability in astronomical spectra. These periodograms, building upon the concept of partial distance correlation, separate the periodic radial velocity modulation induced by orbital motion from that induced by stellar activity. These tools can be used to explore large spectroscopic databases in search of targets in which spectral shape variations obscure the orbital motion; such systems include active planet-hosting stars or binary systems with an intrinsically variable component. We provide a detailed prescription for calculating the periodograms, demonstrate their performance via simulations and real-life case studies, and provide a public Python implementation. 
}


\titlerunning{Partial Distance Correlation Periodograms}
\authorrunning{A. Binnenfeld et al.}

\maketitle
%%%%%%%%%%%%%%%%% BODY OF PAPER %%%%%%%%%%%%%%%%%%
\section{Introduction}
\label{sec:intro}

%{\bf pg1 --- history --- }

The study and interpretation of stellar spectra is a cornerstone of modern astrophysics \citep[e.g.,][]{hearnshaw14}.
During the early days of astronomical spectroscopy, at the end of the nineteenth century, two major fields emerged. 
The discovery of the first spectroscopic binary, $\zeta^1$ Uma (\citealt{pickering1890}, with A.C. Maury; also see \citealt{QUADZETA}), marked the beginning of Doppler spectroscopy, while studies of the relative intensities of spectral lines \citep[][]{cannon1901} laid the foundations for modern stellar classification schemes. 

From a technical point of view, Doppler spectroscopy and stellar classification represent two distinct methodologies, exploiting different parts of the information encapsulated in the observed spectra.
Doppler spectroscopy is focused on the detection of an overall ``shift'' of the spectrum, induced by the relative radial velocity (RV) between the star and the observer \citep[e.g.,][]{tonry79}. Conversely, for the purpose of stellar classification, RV may be considered a nuisance parameter that should be factored out in order to compare the spectral ``shape'' on a common wavelength grid \citep[e.g.,][]{katz98}.

%{\bf pg3 --- important example 1 --- }
In practice, it is not always easy to separate Doppler-shift variations from variations in the spectral shape. For example, line profiles that exhibit time-varying shape deformations can induce errors in Doppler shift measurements, hindering the efforts to detect and characterize extra-solar Earth analogs \citep[see][and references therein]{cameron20}. Even more so, periodic shape variations might mimic exoplanet reflex-motion signatures. In some cases, a careful analysis, correlating RVs with various indicators of stellar activity, suggests that some planet candidates are in fact activity-induced false detections \citep[e.g.,][]{carolo14}.

Classical (type-I) Cepheid variable stars (henceforth: Cepheids) are pulsating yellow (super-)giants that feature large-amplitude spectral line variations due to radial or potentially non-radial pulsation, of up to ${\sim}60\,\,\mathrm{km\,s}^{-1}$ in RV. Their spectra are characterized by phase-dependent significant spectral line asymmetry that complicates the accurate measurement of systematic velocities and introduces biases to the measured pulsational RV amplitudes at the $\mathrm{km\,s}^{-1}$ level \citep[e.g.,][and references therein]{Anderson2018rvs}. Spectra of Cepheids also frequently exhibit wavelength shifts due to binary motion with orbital RV amplitudes up to a couple of tens of $\mathrm{km\,s}^{-1}$, since a majority of Cepheids reside in binary systems \citep{Evans2015rv,Evans2020,Kervella2019a,Pilecki2021}. Distinguishing orbital from pulsational motions, and potentially uncovering the presence of non-radial pulsations, requires a careful distinction between spectral-shape fluctuations and the orbital RV modulations of the binary system \citep[e.g.,][]{Anderson2019polaris}.

The need to analyze complex, composite, and often time-varying spectra led to the development of numerous analysis techniques. For example, studies of double-lined spectroscopic binaries led to the development of TODCOR (a TwO-Dimensional CORrelation technique to analyze stellar spectra) by \citet{zucker94}. Studies of systems that exhibit complex emission features, such as Be stars or accreting compact objects, are analyzed via various distinguishing techniques \citep[e.g.,][]{shenar20}. Advancements in the efficiency of machine-learning algorithms, along with the increase in available computational power, enable new approaches to the analysis of stellar spectra \citep[e.g.,][]{beurs2020}.

This work presents a method to detect periodic modulations in a time series of spectroscopic observations that feature ``shift'' or ``shape'' modulations, or both. The novelty of the presented method is its ability to separate and distinguish periodic signals that originate from Doppler shift from those that originate from line deformation. It is an extension of the phase distance correlation (PDC) concept \citep{zucker18} and the recently introduced unit-sphere representation periodogram (USuRPER) presented in \citet{binnenfeld20}.

The structure of the paper is as follows: Section~\ref{sec:Stat} presents the statistical framework and introduces a detailed prescription for the periodogram calculation. Section~\ref{sec:ex1} demonstrates the application of the new periodograms to simulated test cases, and Section~\ref{sec:ex2} presents the analysis of two classical Cepheids. We summarize our findings in Section~\ref{sec:conc} and discuss the method and its potential for future study.

\section{Statistical framework}
\label{sec:Stat}

Distance covariance, $\mathcal{V}^2$, is a measure of statistical dependence between two random variables introduced by \citet*{Szeetal2007}. Normalizing the distance covariance gives rise to distance correlation, similarly to the way Pearson correlation is derived from the covariance.
Unlike the usual Pearson correlation statistic, $\mathcal{V}^2$ is zero if and only if two random variables are statistically independent. This property of distance correlation was utilized to formulate the PDC periodogram for the purpose of periodicity detection in astronomical time series \citep{zucker18,zucker19}. 
Distance correlation also provides a measure of dependence between random variables of different dimensions. Building on this attribute, \citet{binnenfeld20} extended the use of distance correlation, by introducing USuRPER---a novel method to detect periodic variation in a time series of astronomical spectra. The novelty of USuRPER is that it enables the detection of periodic variations of the overall spectral shape, without extracting specific scalar quantities such as RV or activity indicators from the spectrum. 

This work expands the previous developments, by incorporating the concept of partial distance correlation \citep{szeriz2013} in order to separately identify periodic Doppler and spectral-shape modulations. 

\subsection{Partial and semi-partial correlation} 

In the following introductory subsection we briefly describe the partial and semi-partial correlation, which are designed to control the effect of nuisance parameters on the result of linear regression.

We can consider, as a simple example, a hypothetical experiment that is conducted in order to assess the relation between stellar colors and spectroscopic surface temperatures.
In order to perform this experiment, we select a sample of stars. For each star in the sample we measure color, effective temperature and line-of-sight dust column density. Denote these measured quantities by $(x_i, y_i, z_i)$, respectively, where $i$ refers to the $i$-th star.
In order to simplify the notation, subtract the mean value from each sample and define
\begin{equation}
    \label{eq: minus_mean}
    \vec{x}  = 
    \begin{pmatrix}
    x_1-\overline{x} \\
    \vdots\\
    x_n -\overline{x} 
    \end{pmatrix} \,,  \quad 
    \vec{y} = 
    \begin{pmatrix}
    y_1-\overline{y} \\
    \vdots\\
    y_n -\overline{y}
    \end{pmatrix}    \,,  \quad 
    \vec{z} = 
    \begin{pmatrix}
    z_1-\overline{z} \\
    \vdots\\
    z_n -\overline{z} 
    \end{pmatrix} \, ,
\end{equation}
where $n$ is the number of stars in the sample. 

We assume that the measured values are drawn from a joint distribution of three random variables, ${X}$, ${Y}$ and ${Z}$, which cannot be assumed to be independent. On the one hand, dust affects the color of the observed stars, and therefore ${X}$ may be correlated with ${Z}$. On the other hand, dust is abundant toward the galactic center where late-type stars are common, so ${Y}$ may be correlated with ${Z}$ as well.
The covariance between the random variables can be identified with the standard Euclidean inner product, $\braket{\cdot, \cdot }$. For example,
\begin{equation}
    \label{eq: cov}
    \braket{\,\vec{x}, \vec{z}\,} =   \sum_i (x_i - \overline{x})(z_i - \overline{z})  =  n \,{\rm Cov}(X, Z) \, .
\end{equation}

One way to control the effect of the variable ${Z}$ on the relation between ${X}$ and ${Y}$, is to separately fit two linear relations: One between ${X}$ and ${Z}$, and the other between ${Y}$ and ${Z}$. The residuals from these fits, denoted $\vec{e}_{x}$ and $\vec{e}_{y}$  respectively, are presumably less affected by ${Z}$. Once obtained, the correlation coefficient between $\vec{e}_{x}$ and $\vec{e}_{y}$ can be calculated, and provide a measure of correlation while controlling the effect of the nuisance parameter $Z$. Karl Pearson, who first introduced this approach, coined the term ``partial correlation'' to describe it \citep{Pea1915}.

According to Eq.~\ref{eq: cov}, the residual vectors with respect to $Z$ are given by
\begin{equation}
    \label{eq: ev}
    \vec{{e}_{w}} = \vec{w} - \frac{\braket{\vec{w} , \ \vec{z}}}{\braket{\vec{z} , \vec{z}}}\vec{z} \, .
\end{equation}
For example, in order to obtain the residuals of ${X}$ with respect to $Z$, replace $\vec{w}$ with $\vec{x}$ in Eq.~\ref{eq: ev}.
The partial correlation is the correlation coefficient of the regression residuals. Since the residuals of a regression procedure are of zero mean, the correlation coefficient is simply
\begin{equation}
    \label{eq: partcorr1}
    \mathcal{R}_{xy} = \frac{\braket{\vec{{e}_{x}}, \vec{{e}_{y}}}}{\sqrt{\braket{\vec{{e}_{x}}, \vec{{e}_{x}}}\braket{\vec{{e}_{y}}, \vec{{e}_{y}}}}} \,. 
\end{equation}

By analogy, the example above can be thought of in terms of a projection onto a subspace orthogonal to $Z$.
The correlation coefficient $R_{xy}$ can then be thought of as the cosine of the angle between the vectors $\vec{e}_{x}$ and $\vec{e}_{y}$.

As explained above, partial correlation is designed to control the effect of $Z$ on the two other variables, $X$ and $Y$. However in some cases one may be interested in controlling the effect of $Z$ only on one of the two variables, say, $X$. Such a procedure is called ``semi-partial correlation'', and it is calculated by
\begin{equation}
    \label{eq: partcorr2}
    R_{xy} = \frac{\braket{\vec{e}_{x}, \vec{y}}}{\sqrt{\braket{\vec{e}_{x}, \vec{e}_{x}}\braket{\vec{y}, \vec{y}}}} \,. 
\end{equation}
Building on the heuristic example above, assume that the temperature estimate is not affected by dust. In this case, $Y$ is independent of $Z$. An attempt to control the contribution of $Z$ to $Y$ may lead to over-fitting of the data. Since we still wish to control the effect of dust on the estimated stellar color, we derive the correlation according to equation~(\ref{eq: partcorr2}).

%
% ------------------------------------------------ 
\subsection{Semi-partial distance correlation}
A generalization of partial and semi-partial correlation can be achieved by extending the Euclidean analogy above to some other abstract Hilbert Space.
In the following subsection, we introduce the Hilbert space of $\mathcal{U}$-centered matrices, based on which the semi-partial distance correlation is defined \citep{szeriz2013}.

We can consider a symmetric square matrix in which each component represents the distance between two samples of the same random variable. For example, the distance matrix of the $X$ variable, for a given metric function $d_X$ defined on $X$, is
\begin{equation}
\label{dis}
    x_{ij} \equiv d_X(x_i, x_j) \, ,
\end{equation}
where $i$ and $j$ are the measurement indices.

In analogy to the mean subtraction in Eq.~\ref{eq: minus_mean}, $\mathcal{U}$-centering is applied to the distance matrix, producing a new matrix $\mtrx{x}$ \citep{szeriz2013}. 
The prescription for the elements of the $\mathcal{U}$-centered matrix is 
\begin{equation}
{\mtrx{x}}_{\, ij} = 
x_{ij} - \overline{x}_{ij} ,
\end{equation}
where
\begin{equation}
    \overline{x}_{ij} = 
    \begin{cases}
        \frac{1}{n-2}\bigg[\sum_{k}(x_{ik} + x_{kj})
        - \frac{1}{(n-1)}\sum_{kl}x_{kl}\bigg], & \text{if } i\neq j \, ,\\ \\
        x_{ij} &\text{otherwise.}
    \end{cases}
\end{equation}

The $\mathcal{U}$-centered matrices form a Hilbert space together with the inner product 
\begin{equation}
    \label{eq: inprod}
    \braket{\mtrx{x}, \mtrx{z}}_{_\mathcal{U}} \equiv \frac{1}{n(n-3)}\sum_{i\neq j}{\mtrx{x}_{ij}\mtrx{z}_{ij}} \, .
\end{equation}
\citet{szeriz2013} have shown that this expression is an unbiased estimator of the distance covariance, $\mathcal{V}^2$. 
By relating the inner product with the distance correlation, we have endowed the abstract geometrical properties of the space with a powerful statistical meaning. Now, the semi-partial distance correlation can be defined in an identical manner to the definitions provided in equations~(\ref{eq: ev}) and (\ref{eq: partcorr2}),
\begin{equation}
    \label{eq: ten}
    \mtrx{e}_{w} = \mtrx{w} - \frac{\braket{\mtrx{w} , \ \mtrx{z}}}{\braket{\mtrx{z} , \mtrx{z}}}\mtrx{z} \, ,
\end{equation}
where $\mtrx{w}$ stands for the $\mathcal{U}$-centered distance matrix of the $X$ or $Y$ random variable. The semi-partial distance correlation between $X$ and $Y$, controlling for the effect of the nuisance variable $Z$ on $X$, is therefore
\begin{equation}
    \label{eq: partcorr3}
    D_{xy} = \frac{\braket{\mtrx{{e}_{x}}, \mtrx{y}}_{_{\mathcal{U}}}}{\sqrt{\braket{\mtrx{{e}_{x}}, \mtrx{{e}_{x}}}_{_{\mathcal{U}}}\braket{\mtrx{y}, \mtrx{y}}}_{_{\mathcal{U}}}} \,.
\end{equation}

\subsection{Partial distance correlation periodograms}
\label{partial}
We can consider a time series of spectroscopic observations, and assume that the spectrum is changing with some periodicity $P$. 
We associate each measurement epoch, $t_i$, with its corresponding phase, $\phi_i\equiv t_i\mod P$, given in units of time. 
As described in equation~(\ref{dis}), a distance matrix $\mtrx{\phi}$ is computed based on the phase metric, $d_{\Phi}$, given by
\begin{equation}
\label{eq: phase metric}
% \begin{split}
 d_{\Phi} = \phi_{ij}\cdot(P-\phi_{ij}) \ ,
% \end{split}
\end{equation}
where $\phi_{ij} = (t_i - t_j)\mod P$ \citep[see][]{zucker18}.
For each spectrum we estimate the RV, $v_i$, and compute its distance matrix, $\mtrx{v}$, using the RV metric, 
\begin{equation}
\label{eq: rv metric}
d_v = |v_i - v_j| \ .
\end{equation}
The phase and RV construct the first two random variables in the problem. The third is the spectral shape, $s_i$. The $\mtrx{s}$ distance matrix is computed as defined by \citet{binnenfeld20}, using the metric
\begin{equation}
d_s = \sqrt{1-C_{ij}} \, ,
\label{eq:C}
\end{equation}
with $C_{ij}$ being the correlation between a pair of spectra, $i$ and $j$, shifted the naively assumed rest frame according to $v_i$. This correlation is normalized by subtracting the spectrum mean value and dividing it by its standard deviation. 

To construct a periodogram in which the influence of spectral-shape variability is controlled, we calculate  $D_{{{\phi}v}}$, the semi-partial distance correlation between the phase and the RV, according to equations (\ref{eq: phase metric}) and (\ref{eq: rv metric}) while treating the spectral shape, from equation~(\ref{eq:C}), as a nuisance parameter. The calculation is repeated for an array of trial periods, resulting with a ``shift'' periodogram, which is henceforth abbreviated $D_{{v}}$.
Conversely, when using the RV as a nuisance variable, the periodogram discards any signal related purely to Doppler-shift variability, accounting only for periodic shape variability $D_{{{\phi}s}}$. As before, by calculating the Doppler-shift-controlled semi-partial distance correlation for an array of trial periods, we construct a ``shape'' periodogram, which is henceforth abbreviated $D_{{s}}$. A summary of our notation is presented in Table~\ref{KapSou}.

   \begin{table*}
      \caption[]{Summary of periodogram notation conventions.}
         \label{KapSou}
     $$ 
         \begin{array}{lcll}
            \hline
            \noalign{\smallskip}
            \textrm{Periodogram} & \textrm{Notation} & \textrm{Sensitive to} & \textrm{Nuisance variable} \\
            \noalign{\smallskip}
            \hline
            \noalign{\smallskip}
            \textrm{partial PDC} & D_{{v}} & \textrm{Doppler ``shift''}  & \textrm{non-Doppler ``shape'' variability}   \\
            \textrm{partial USuRPER} & D_{{s}} & \textrm{non-Doppler ``shape'' variability} & \textrm{Doppler ``shift''} \\
            \noalign{\smallskip}
            \hline
         \end{array}
     $$ 
   \end{table*}

We chose to use semi-partial distance correlation, according to the prescription in equation~(\ref{eq: partcorr3}), rather than partial distance correlation, since the semi-partial approach yielded better results. Therefore, in practice,  we control the mutual dependence between Doppler shift and spectral shape, without eliminating the chosen nuisance variable dependence on the phase.

Stellar activity and pulsation usually manifest in general changes in spectral shape and spectral line deformation, while orbital motion indicating the existence of a stellar or substellar companion will naturally result in Doppler-shift variability. Therefore, our novel periodogram can be used to distinguish stellar activity and pulsation from orbital RV shifts. This can help detect planets orbiting active stars whose activity once obscured the planetary signal, identify activity cycles mimicking planetary signals, and characterize observed multiple systems including an active or pulsating component. 

Following \citet{binnenfeld20}, we assess the peak significance in those periodograms using permutation tests. We allocate random phases (drawn uniformly) to each spectrum, and recalculate $D$ for this random allocation of phases. We create a sample of $D$ values by repeating the shuffle, and use it to obtain a threshold value matching a desired level of false-alarm probability (FAP).

\section{Simulations}
\label{sec:ex1}
After introducing the conceptual and statistical framework of the partial distance correlation periodograms, we now demonstrate their ability to distinguish periodic shift from shape modulations via simple numerical experiments. The numerical experiments present two limiting cases of shape variability: a strictly periodic shape modulation,  and a completely random one.

One way to induce spectral shape variability is stellar activity. That, for example, when spots appear, disappear, or change position on the surface of the star. The resulting spectral features periodically modulate with the stellar rotation, which varies at different latitudes \citep[e.g.,][]{Heitzmann21}. For the purpose of the experiments, we generate sets of synthetic spectra, which we shift as if reflex motion is induced by a planet and deform by injecting spectral features that presumably stem from the presence of stellar spots. To simulate the observed spectra, we used PHOENIX spectral library \citep{Husetal2013}, assuming an effective stellar temperature of $5800$ K, $\log g$ of $4.5$ and a Solar metallicity. The spectra were sampled in the wavelength range of $5920$\,--\,$6000$\,\AA, and broadened assuming a rotational velocity of  $v\,\sin{i} = 5\,\mathrm{km\,s}^{-1}$. 

A simplistic stellar spot model was used to generate the shape modulations. We assume that spots can be regarded as relatively cold and dark patches on the surface of the star. Therefore, to simulate their effect, we subtracted a copy of the non-broadened stellar spectrum, weighted according to the assumed spot size and shifted to the instantaneous velocity of the spot. A spectral template of a colder star, of $2300$ K, was inserted in its stead, with the same shift and assuming the same spot size. Doppler shift was then introduced to the deformed spectrum, according to an assumed set of orbital elements of a hypothetical planet. Finally, the spectrum was broadened with a Gaussian profile, to simulate an instrumental resolution of $R=100000$, and white noise was added to the data.

We analyze the simulated data sets with our newly introduced periodograms, and compare our findings with results we obtain with other, well-established, methods. This includes correlating the bisector inverse span (BIS) and full-width half maximum (FWHM) of the cross-correlation function (CCF) with the derived RVs \citep[e.g.,][]{Queloz2001, MELL2021}.

\subsection{A simulated periodically active star hosting a planet}

\label{subsec:x8888}

The first simulated test-case is of an active planet-hosting star, where the activity-induced signal is periodic. The challenge in this case stems from the periodic nature of the spot, as its resulting  RV might mimic that of a true planet-induced Doppler shift, resulting in a false planet detection \citep[e.g.,][]{noPlan2004}. Alternatively, the periodic activity may impair the detection efficiency, either by making the data too noisy for Doppler shift to be detected or by making Doppler detection unreliable. 

The star was assumed to have two spots, fixed in latitude and longitude, arranged such that one spot is located on the visible side of the star at all times. The synthetic spectra were deformed accordingly, as described earlier. The frequency of the resulting shape modulation was determined by the assumed 25-day stellar rotation period.

The prominence of the shape modulation was governed by the $1\,\mathrm{km\,s}^{-1}$ RV semi-amplitude of the spot and its assumed $0.5\%$ contribution to the overall flux. After the spectra were deformed, a Doppler shift was induced, shifting the entire bulk of the spectrum, assuming a companion in a moderately eccentric orbit ($e=0.3$). The RV semi-amplitude of the reflex motion was set to $50~\mathrm{m\,s}^{-1}$ and the orbital to $7$ days. The $35$ measurement epochs were randomly drawn from a uniform distribution over an interval of $100$~days. 

\begin{figure*} 
 \centering
\includegraphics[width=1.6\columnwidth,clip=true]{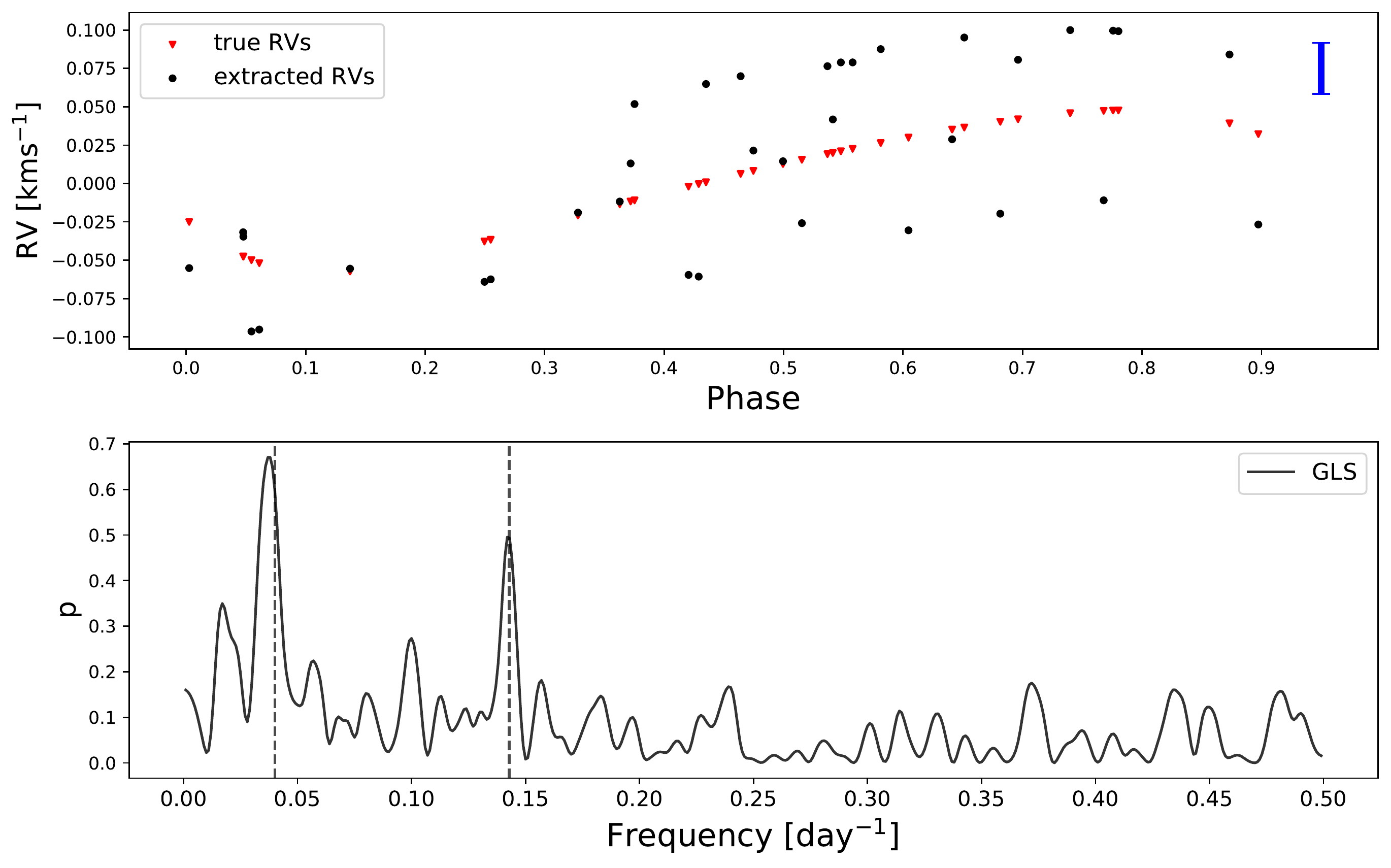}
\caption{RV based analysis for the simulated periodically active planet-hosting star. \textit{Upper panel:} RVs obtained for the spectra through cross-correlating against a PHOENIX template. The RVs are phase-folded according to the planetary period. The blue bar stands for the mean difference between true and extracted RV.
%(in relation to the true RVs). 
\textit{Lower panel:} GLS periodogram for the RV data. The dashed vertical lines mark the expected activity and planetary frequency, matching 7 and 25 days period.}
\label{fig:sim_yes_p_spot_GLS}
\end{figure*}

\begin{figure} 
%  \centering
\includegraphics[width=\columnwidth,clip=true]{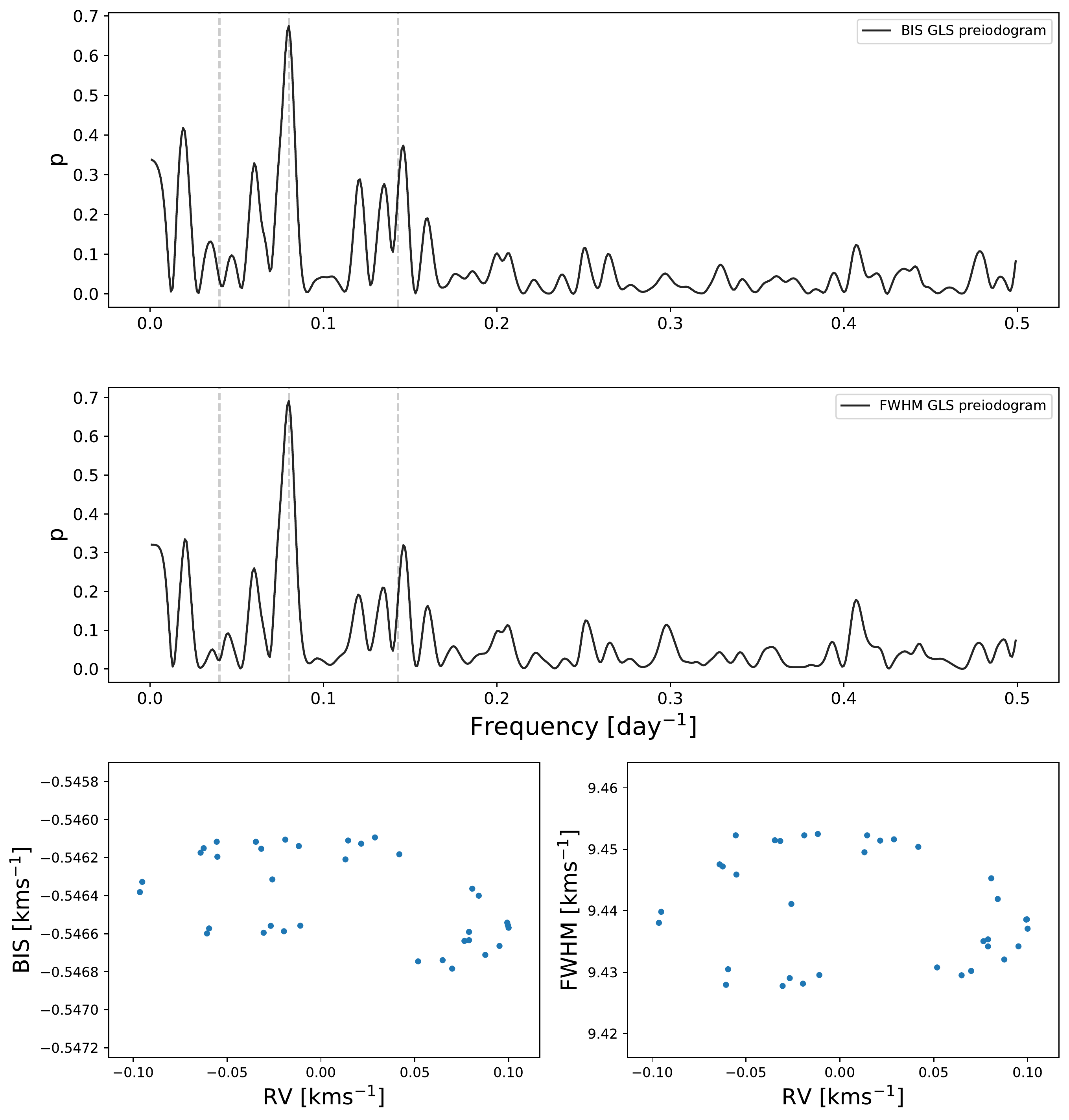}
\caption{Activity-indicators-based analysis of the simulated periodically active star hosting a planet. \textit{Upper and middle panels:} GLS periodogram for the BIS and FWHM data, respectively. The dashed vertical lines mark the expected 7-days planetary frequency, as well as the activity frequency and half-frequency in 25 and 12.5 days period. 
\textit{Lower panels:} Bisector inverse span and full-width half maximum data as a function of RVs obtained for the spectra through cross-correlating against a PHOENIX template.}
\label{fig:activity_2_periods}
\end{figure}

We then analyzed the simulated data set in search of planetary induced periodic RV modulations. The RVs were derived by cross-correlating the spectra with a PHOENIX template, using the \texttt{SPARTA} package.\footnote{Available at \url{https://github.com/SPARTA-dev/SPARTA}} The position of the CCF peak was used to determine the RV, and its BIS and FWHM values were used to identify shape distortions. We searched the derived RV, BIS and FWHM for periodic signals, using the generalized Lomb-Scargle periodogram \citep[GLS;][]{Fer1981,ZecKur2009}, as shown in Fig.~\ref{fig:sim_yes_p_spot_GLS} and Fig.~\ref{fig:activity_2_periods}.
The RV periodogram shows two peaks---one at the simulated orbital shift frequency and another at the spot induced shape frequency. The BIS and FWHM show significant periodic variability of 12.5 days, that is, at double the true activity frequency. 

Spearman correlation tests of the BIS and FWHM against the RVs were performed to determine the nature of the periodic variability (see Table~\ref{spearman_table_1}). 
As shown in Fig.~\ref{fig:activity_2_periods}, the BIS and FWHM are correlated with the RV. However, the Spearman tests yield insignificant results, as the relation is highly nonlinear. At this point, the existence of periodically variable line-deformation is evident, and its effect on the RV is established. Therefore, based on the activity indicators alone it may be challenging to determine which of the detected frequencies, if any, stem from periodic Doppler shift.  

\begin{figure*}
 \centering
\includegraphics[width=1.6\columnwidth,clip=true]{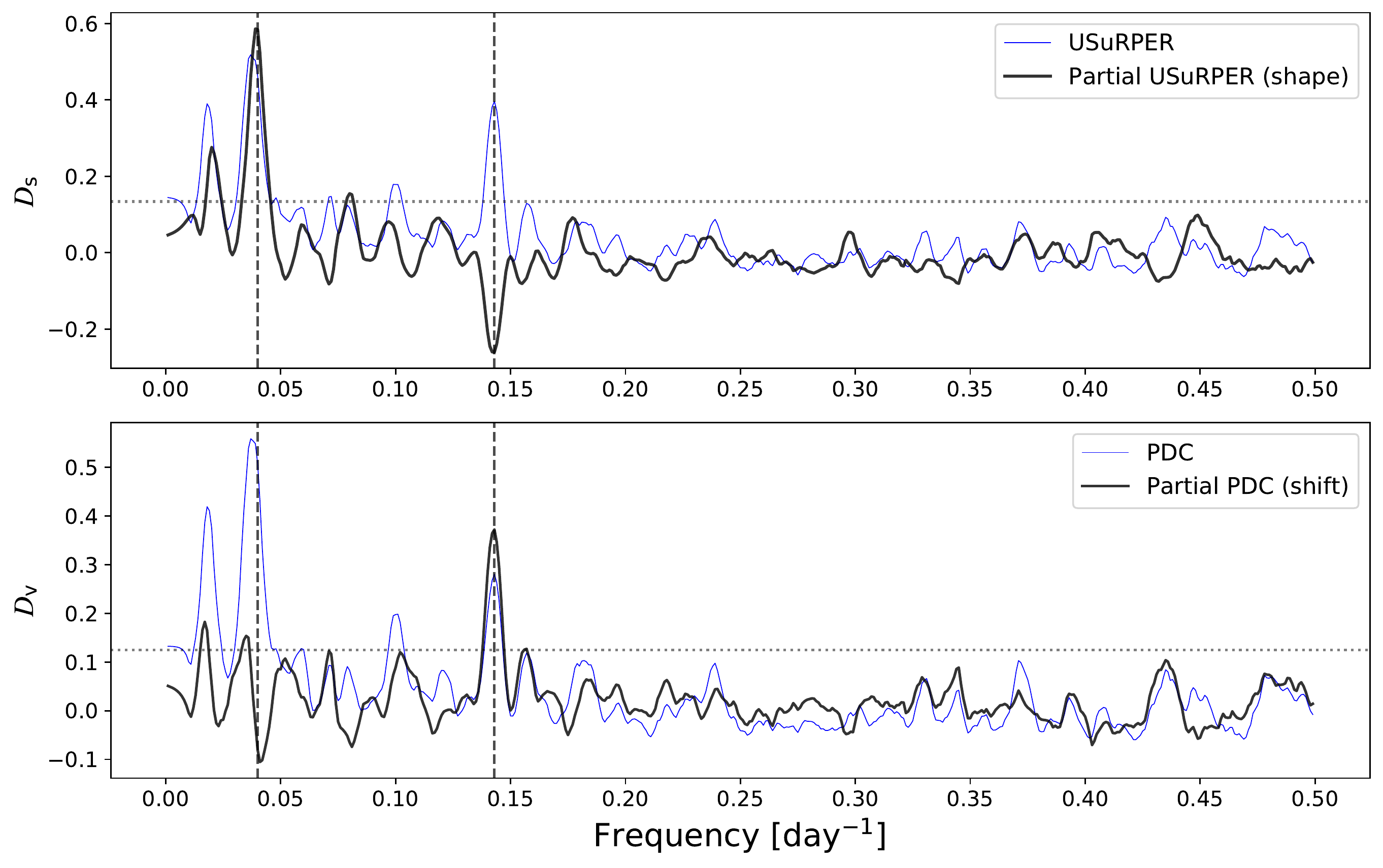}
\caption{Partial periodogram results for the simulated periodically active star hosting a planet. \textit{Upper panel:} USuRPER periodogram results. The dashed vertical line marks the expected activity and planetary frequencies ($25$ and $7$ days). USuRPER shows significant peaks in both frequencies, while the partial USuRPER, the shape periodogram, shows a significant peak in the activity frequency only. \textit{Lower panel:} Matching PDC periodograms. As it is designed to do, the partial PDC, the shift periodogram, presents only the peak matching the planetary period (7 days), and not the one matching the activity period as the PDC periodogram. The dotted horizontal line in both panels corresponds to a FAP level of $10^{-4}$, obtained by the permutation test procedure.}
\label{fig:sim_yes_p_spot}
\end{figure*}

Similar to the GLS periodogram, the PDC and USuRPER present peaks in both the activity and the planetary frequencies (blue thin line in Fig.~\ref{fig:sim_yes_p_spot}). The partial periodograms, on the other hand, successfully distinguish the periodic variations, enabling one to identify their sources: the shift periodogram, partial PDC, sensitive to shift modulations, indeed shows a significant peak in the planetary frequency only (thick black line in Fig.~\ref{fig:sim_yes_p_spot} lower panel). The shape periodogram, partial USuRPER, designed to account for the shape modulation, shows a prominent peak in the stellar activity period (thick black line in Fig.~\ref{fig:sim_yes_p_spot} upper panel).

   \begin{table}
      \caption[]{Spearman correlation coefficients between the RVs and the BIS and FWHM activity indicators.}
         \label{spearman_table_1}
     $$ 
         \begin{array}{llrr}
            \hline
            \noalign{\smallskip}
             \multicolumn{1}{l}{\textrm{Case}} & 
             \multicolumn{1}{l}{\textrm{indicator}} & 
             \multicolumn{1}{c}{r_s}  &
             \multicolumn{1}{c}{\textrm{p-value}} \\
            \noalign{\smallskip}
            \hline
            \noalign{\smallskip}
            \textrm{I} &  \textrm{BIS}  & \textrm{-0.13}  & {0.44} \\
            \textrm{I} &   \textrm{FWHM}  & \textrm{-0.39}  & {0.02} \\
            \noalign{\smallskip}
            \hline
            \noalign{\smallskip}

            \textrm{II} &  \textrm{BIS}  & \textrm{-0.27}  & {0.12} \\
            \textrm{II} &  \textrm{FWHM}  & \textrm{-0.42}  & {0.11} \\ 

            \hline
         \end{array}
     $$ 
        \tablefoot{
        The table presents both simulated cases: Case I---a periodically active star hosting a planet and Case II---a planet hosting star with random spectral variability (see text).}

   \end{table}

%1.3 

%1.3 

% 1.6

\subsection{Planet host with random spectral variability}

Next, we tested the performance of the new periodograms using random, nonperiodic, shape fluctuations. These do not necessarily correspond to modulations of astrophysical origin but do provide useful insights into the behavior of the periodograms in the presence of additional noise sources. In the case of real observations, apparently stochastic fluctuations could be introduced by instrumental instability, for example. Regardless of their origin, such fluctuations can impair the ability to correctly identify periodic Doppler shifts.

\begin{figure*}
 \centering
\includegraphics[width=1.4\columnwidth,clip=true]{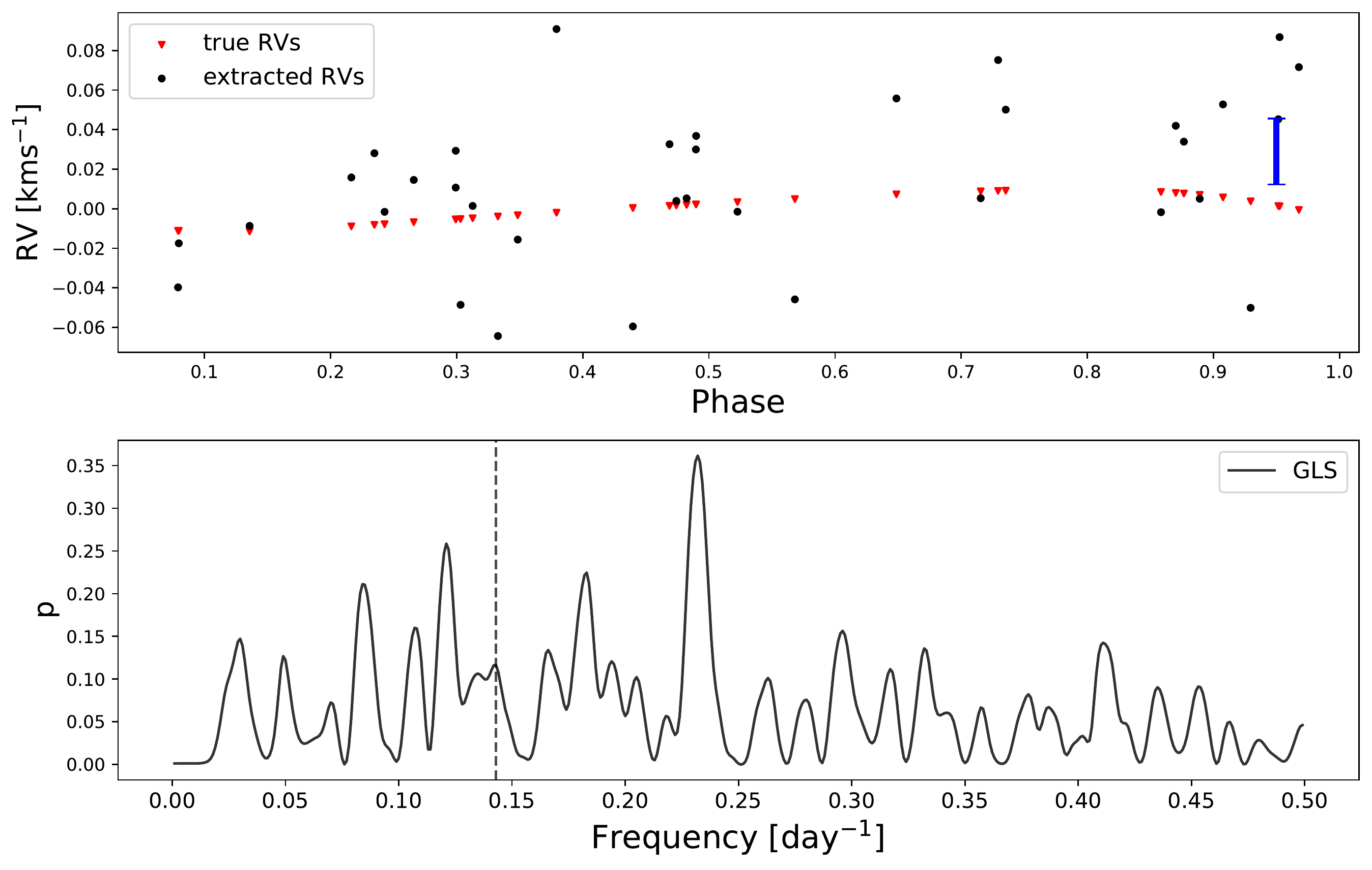}
\caption{RV based analysis of the simulated planet host with random spectral variability. \textit{Upper panel:} RVs obtained from the spectra through cross-correlating against a PHOENIX template. The RVs are phase-folded according to the planetary period. The blue bar stands for the mean difference between true and extracted RV. 
%(in relation to the true RVs). 
\textit{Lower panel:} GLS periodogram for the RV data. The dashed vertical line marks the expected planetary frequency, where no prominent peak is presented due to the spectral variability induced noise.}
\label{fig:sim_no_p_spot_GLS}
\end{figure*}

\begin{figure} 
%  \centering
\includegraphics[width=\columnwidth,clip=true]{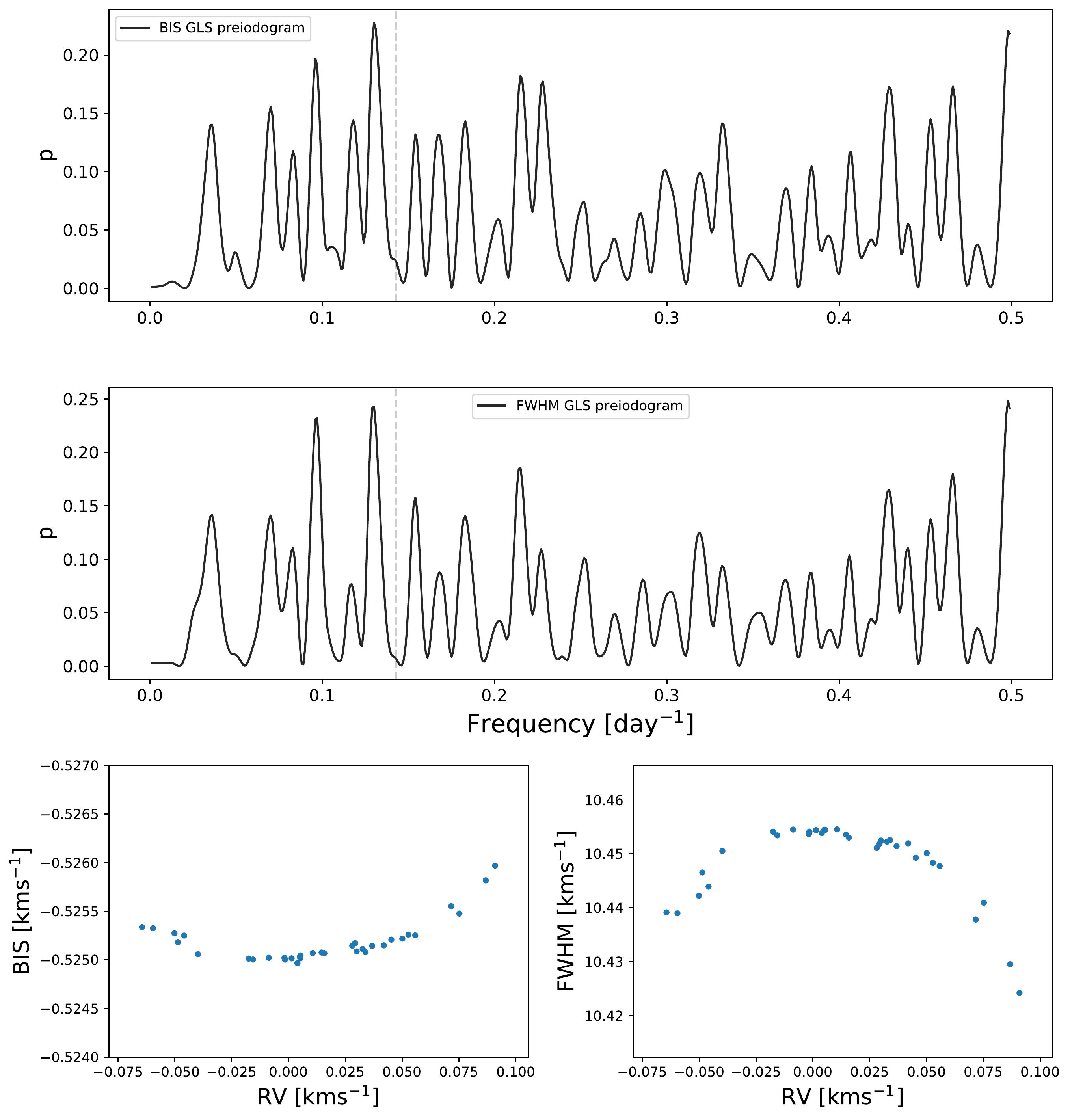}
\caption{Activity-indicators-based analysis of the simulated planet host with random spectral variability. \textit{Upper and middle panels:} GLS periodogram for the BIS and FWHM data, respectively. The dashed vertical lines mark the expected 7-days planetary frequency.
\textit{Lower panels:} Bisector and full-width half maximum data as a function of RVs obtained for the spectra through cross-correlating against a PHOENIX template.}
\label{fig:activity_1_period}
\end{figure}

We simulated nonperiodic shape modulations the same way we simulated the periodic stellar activity in the last section, only this time the underlying RV values were drawn from a uniform distribution in the range $[-2,2]~\mathrm{km\,s}^{-1}$. The planet-induced RV semi-amplitude was set to $10~\mathrm{m\,s}^{-1}$, and the orbit was chosen to be circular. We randomly sampled the system at 35 epochs. Similarly to the previous test case, the parameters of the simulation were chosen arbitrarily for demonstration purposes.
Fig.~\ref{fig:sim_no_p_spot_GLS} shows the resulting RV estimates, as well as the matching GLS periodogram. The simulated spectra generated a noisy RV time series, in which the seven-day planetary sinusoidal periodicity is obscured by spectral variability. The matching GLS periodogram is noisy as well, and shows no peak in the planetary frequency.
As demonstrated in Fig.~\ref{fig:activity_1_period}, both activity indicators are correlated with the RV. However, as in the previous simulation, the Spearman test does not yield a significant p-value (see Table~\ref{spearman_table_1}), as the relation is highly nonlinear. Therefore, even if one detects the 7-day signal, redeeming it as being Doppler- rather than activity-induced, will require further analysis. The numerical difference between the simulated test cases resulted in a somewhat different correlation pattern of the activity indicators and the RVs. 
%Note that the difference in the correlation pattern of the activity indicators and the RVs between the two simulated test cases can be explained by their numerical differences, a fact established using additional simulated cases tested.
%

\begin{figure*}
\centering
\includegraphics[width=1.4\columnwidth,clip=true]{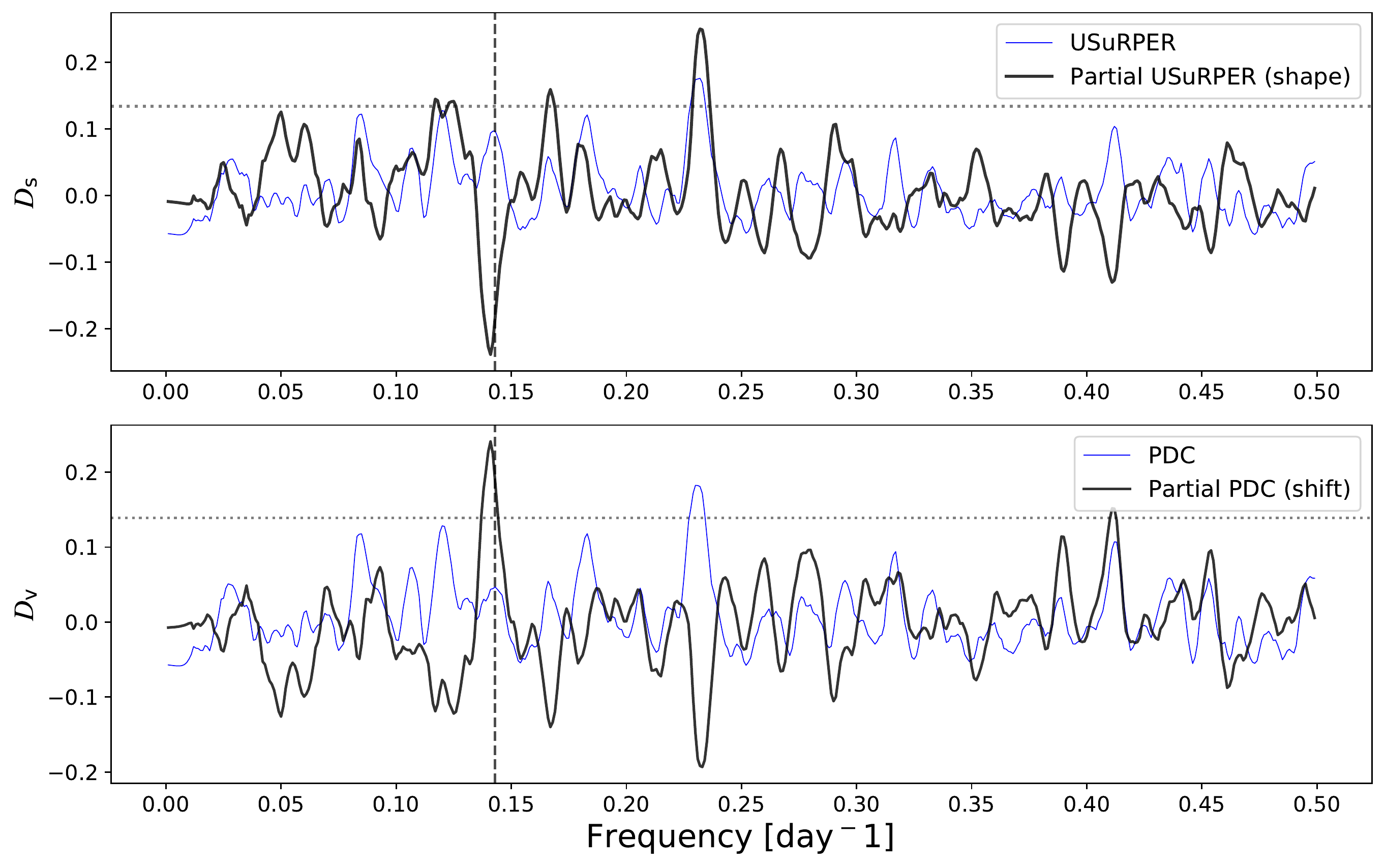}
\caption{Partial periodogram results for the simulated planet host with random spectral variability. \textit{Upper panel:} USuRPER and partial USuRPER (shape) periodogram results. The dashed vertical line marks the expected planetary frequency, in which no significant peak is present. \textit{Lower panel:} PDC and partial PDC (shift) periodogram results for the same data. No peak appears in the PDC periodogram, as the planetary signal is obscured by the spectral shape variations induced noise. In the matching shift periodogram (partial PDC), on the contrary, a significant peak does appear in the expected planetary frequency. The dotted horizontal line in both panels corresponds to a FAP level of $10^{-4}$, obtained by the permutation test procedure.}
\label{fig:sim_no_p_spot}
\end{figure*}

Fig.~\ref{fig:sim_no_p_spot} presents the PDC and USuRPER periodograms for the simulation, neither of which exhibit a response at the expected planetary frequency, nor does the partial-USuRPER shape periodogram. Nevertheless, the partial-PDC shift periodogram presents a prominent peak at the planetary frequency.

% 1.3 TBD

%1.3 

%1.6

\section{Pulsating stars}
\label{sec:ex2}

In the previous section, we applied our newly introduced methods on sequences of simulated spectra, which are far better behaved than real-life data. We therefore tested our periodograms on real-life spectra of two classical Cepheid variables, one of which is part of a spectroscopic binary system. Their pulsation, like the stellar activity in the simulated cases, leads to significant modulations of spectral shape, rendering these objects appropriate test cases for the new methodology.

\subsection{$\beta$ Doradus -- HD\,37350}

$\beta$~Dor is a naked-eye ($V \simeq 3.8$\,mag) classical Cepheid in the southern hemisphere. Its proximity to the Southern Ecliptic Pole places $\beta$~Dor within the \textit{TESS} continuous viewing zone and allows a very detailed study of its photometric variability, whose period is ${\sim}9.84318$\,d \citep{TESS2020}. $\beta$~Dor 
has not previously shown any indications of having a spectroscopic companion, making this a suitable test case for the separability of positional and shape modulations. This is particularly interesting, since radial pulsations simultaneously modulate line-of-sight velocity and spectral line shape with the same periodicity. With a period near 10 days, $\beta$~Dor's RV curve is characterized by a significant bump feature that can be closely reproduced by a 7-harmonic Fourier series fit to the RV data.

We analyzed $135$ spectra observed as part of the GE-CeRVS project \citep[e.g.,][Anderson et al. in prep.]{Anderson2019rvs} with the high-resolution optical echelle spectrograph CORALIE ($R\simeq60000$) on the Swiss $1.2$ meter Euler telescope at La Silla Observatory, Chile. Data reduction was performed using the dedicated CORALIE data reduction pipeline at the University of Geneva. This pipeline performs all required steps, including bias and flatfield correction, cosmic ray rejection, and used \element{Th}-\element{Ar} lamps for wavelength calibration. The typical signal-to-noise ratio (S/N) of the spectra is approximately $110$ at $5000$\AA. We restricted our analysis, somewhat arbitrarily, to a narrow wavelength range of $4900$\,--\,$5150$\,\AA. We tested our method using additional sections of the spectrum, yielding comparable results.

\begin{figure*}
 \centering
\includegraphics[width=1.6\columnwidth,clip=true]{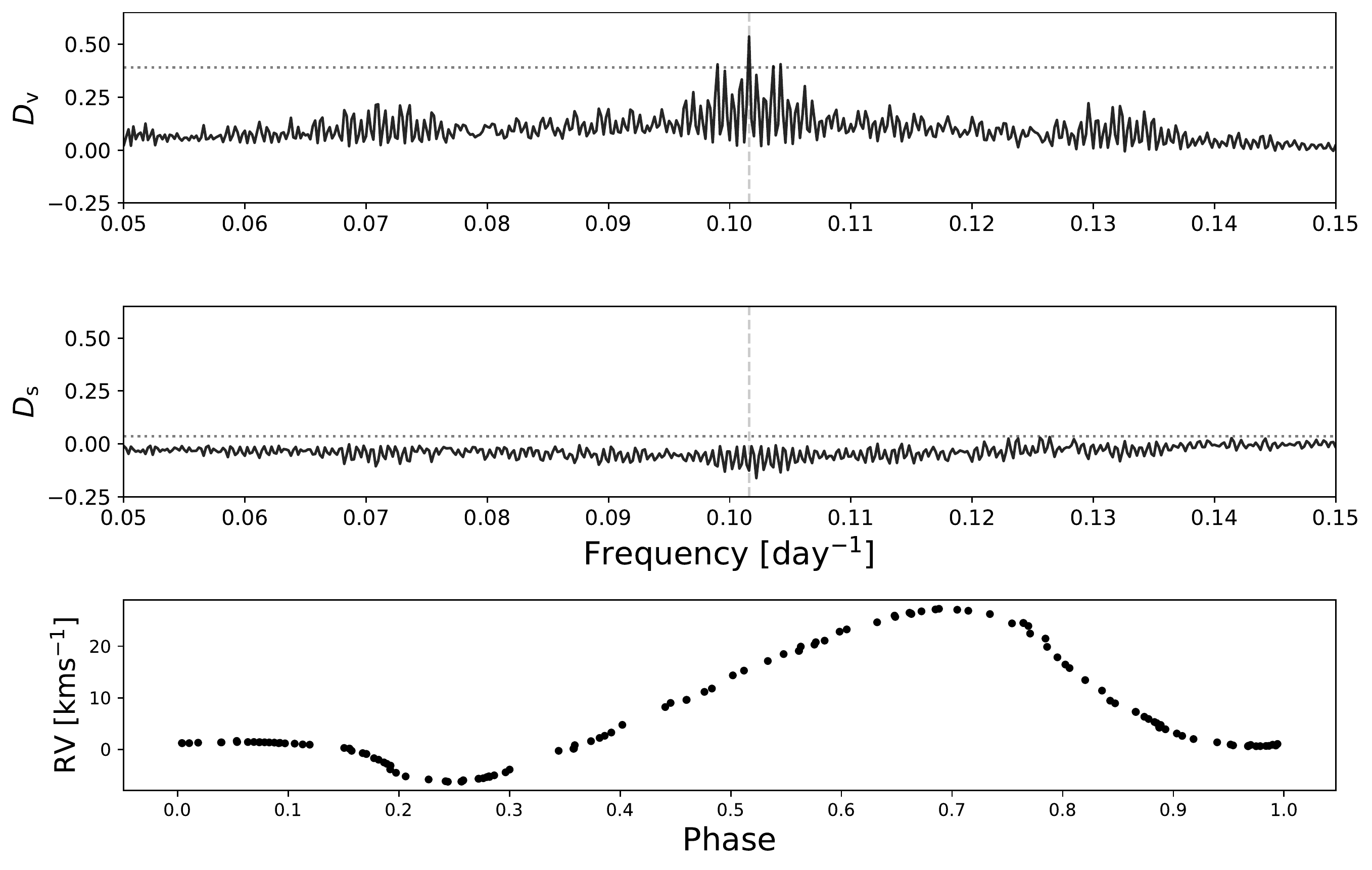}
\caption{Partial periodogram results for $\beta$~Doradus. \textit{Upper panel:} Shape periodogram (partial USuRPER) result, presenting a clear peak at the 9.84318 days signal due to pulsation.
\textit{Middle panel:} Shift periodogram (partial PDC) results eliminating the peak, as it is designed to be sensitive only to positional modulations, such as pure Doppler shifts due to orbital motion.
\textit{Lower panel:} RVs extracted for the observations, folded according to the 9.84318 days pulsation period. The dotted horizontal line in both panels corresponds to a FAP level of $10^{-4}$, obtained by the permutation test procedure.}
\label{fig:betador}
\end{figure*}

Fig.~\ref{fig:betador} illustrates the results obtained and demonstrates the ability of the new approach to recover the known $9.84318$\,d periodicity in the RV time series. The significant periodicity also manifests in both the PDC periodogram and the USuRPER. However, the shift periodogram, that is, the method sensitive only to displacements in line position, shows no obvious peak in the variation frequency. Effectively, using this approach one can safely attribute all the RV variability to variability of the spectrum shape due to the Cepheid pulsations, and not to any periodic bulk motion.

% 1.6

\subsection{S Muscae -- CD-69\,977}
\label{subsec:SMUS}

S~Mus is a bright, $V \simeq 8.3$ mag, southern spectroscopic binary Cepheid with $505$\,d orbital and $9.65996$\,d pulsational periods \citep[e.g.,][]{EVA1990,Gallenne2019}. Its pulsational RV variability is very similar to that of $\beta$~Dor owing to the similar period. The main differences between the two cases are the number of observations, the time span covered, the S/N of the observations, and the additional signal due to orbital motion.

Given the two types of signals present in the spectra, we expect the shape periodogram to exhibit a significant peak at the known pulsation frequency and the shift periodogram to exhibit a significant peak at the known orbital frequency. As was the case for $\beta$~Dor, the Doppler shift modulation associated with pulsation is expected to be absorbed into the shape modulation at the identical frequency.

\begin{figure*}
 \centering
\includegraphics[width=1.6\columnwidth,clip=true]{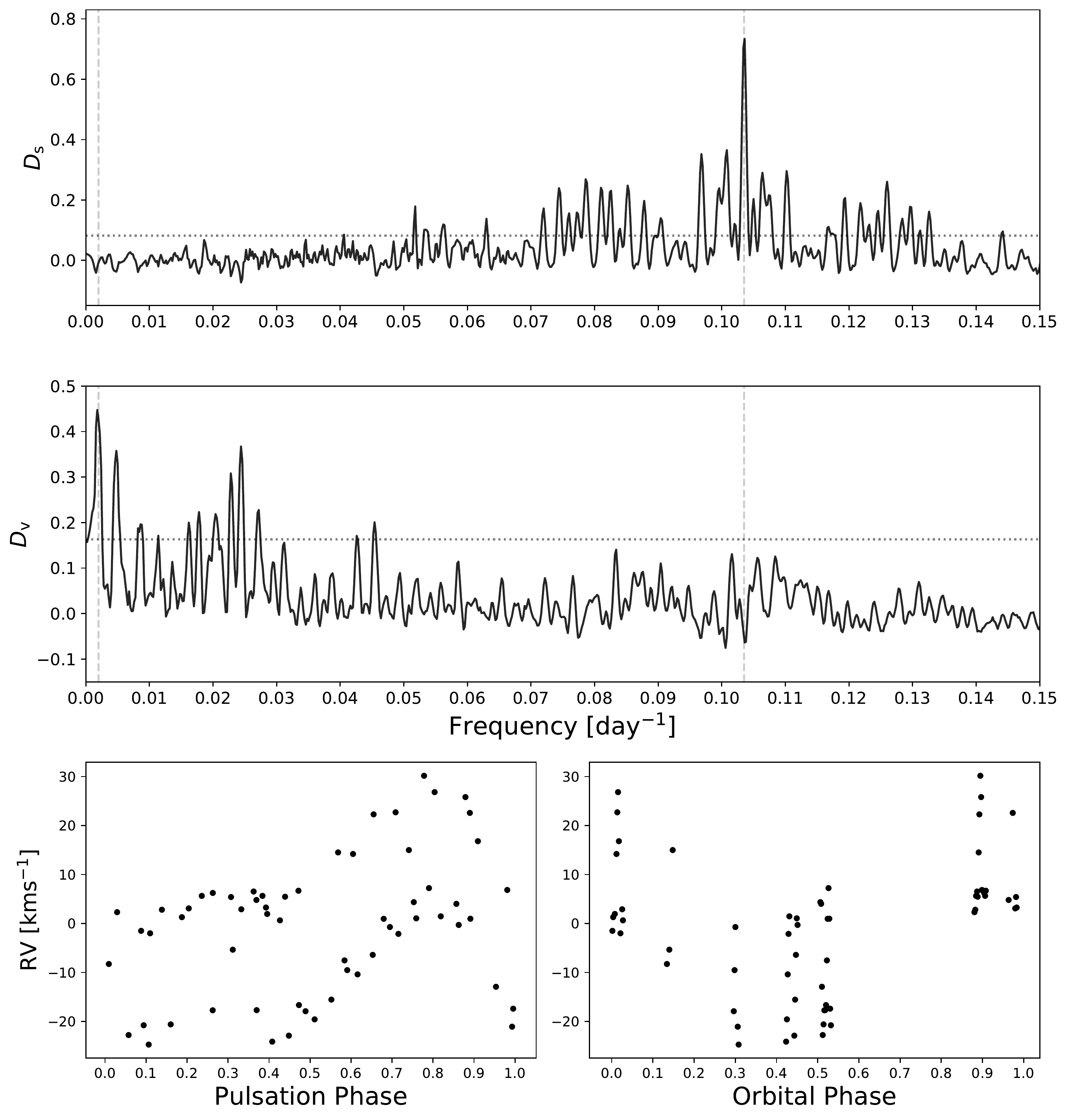}
\caption{Partial periodogram results for S Muscae. \textit{Upper panel:} Shape periodogram (partial USuRPER) result. The vertical dashed lines represent the 505 days orbital period and the 9.65996 days pulsation period. The shape periodogram is designed to account for spectral shape variation signals only, thus eliminating the 505 days signal originated in orbital motion.
\textit{Middle panel:} Shift periodogram (partial PDC) result, designed to account for Doppler-shift variations only, therefore showing a significant peak in the 505 days orbital period, while eliminating the 9.65996 days pulsation originated peak. An additional peak, matching a 40 days period, appears in the periodogram and might be caused by some kind of a window function related to the observation times.
\textit{Lower panels:} RVs extracted for the observations, folded according to the 9.65996 days pulsation phase (\textit{left}) and 505 orbital phase (\textit{right}). The dotted horizontal line in both panels corresponds to a FAP level of $10^{-4}$, obtained by the permutation test procedure.}
\label{fig:SMUS}
\end{figure*}

We collected $60$ CORALIE spectra of S~Muscae, observed between January 2012 and May 2018, and computed the periodograms on the somewhat-arbitrary $5600$\,--\,$5800$\,\AA \,wavelength range. As shown in Fig.~\ref{fig:SMUS}, the partial periodograms successfully distinguish the two periodic variations: the shift periodogram shows a clear significant peak in the $505$ days orbital frequency, while the shape periodogram reveals the underlying spectral shape fluctuation in the form of a prominent peak in the $9.65996$-days pulsation period. The second highest peak in the shift (partial PDC) periodogram near ${\sim}40$\,d is most likely related to sampling, that is, the window function.

% 1.6

\section{Summary and conclusions}
\label{sec:conc}

In this work, we introduced the partial distance correlation-based periodograms, which are able to distinguish Doppler shift from spectral shape variability in astronomical spectra. 

We tested the periodogram performance in simulations, focusing on two limiting cases: strictly periodic and completely random non-Doppler spectral variability. These examples attest to the broad potential of the periodograms, which were shown to be effective in distinguishing periodic modulations of line position and shape in cases where their time scales are well separated. We successfully tested our method on real data as well, using observational data of two classical Cepheids, one of which is a spectroscopic binary.

One noteworthy property of these new periodograms is the existence of negative values. This feature originates from the partial distance correlation unbiased estimator, given in equation~(\ref{eq: partcorr3}), which, as an inner product, may attain both negative and positive values. This phenomenon can be seen in Figs.~\ref{fig:sim_yes_p_spot} and \ref{fig:sim_no_p_spot} which present the periodogram for the simulated cases, and we expect it can be encountered in some real-life cases as well. Unlike standard correlation coefficients, ``positive'' partial distance correlation values imply statistical dependence, but ``negative'' values do not necessarily carry a simple, opposite, meaning. For example, a case where the controlled nuisance parameter carries significant phase dependence can be considered. In such a case, a peak in one partial periodogram can be accompanied by a negative one (a ``trough'') in the other, because removing the contribution of that dependence may result in negative values of relatively large magnitudes, as the numerator of equation~(\ref{eq: partcorr3}) suggests.

In cases where the Doppler shift is strongly associated with line shape deformation, for example, radial pulsation of a classical Cepheid, periodic shape and shift modulations will generally remain indistinguishable. This is a property of the algorithm that stems from the fact that the shape metric in equation~(\ref{eq:C}) is calculated after the spectra are shifted according to the derived RVs. To investigate this property, we experimented with simulated cases where a Doppler shift and stellar activity vary with the same frequency. We found that only the shape periodogram exhibited a significant peak, while the shift periodogram did not.

It is especially encouraging to see this quality of the method demonstrated in both Cepheid cases we examined. The shift periodogram was insensitive to the pulsational Doppler-shift variability caused by the stellar radius changes, despite its large magnitude of a few tens of $\mathrm{km\,s}^{-1}$, presented in the bottom panels of Fig.~\ref{fig:betador} and Fig.~\ref{fig:SMUS}  \citep[see][]{Anderson2018rvs}. This shows that the periodogram indeed distinguishes the periodic-line shape distortions from the Doppler shift, and therefore likely to help prevent false planet detections due to periodic line shape variations. Furthermore, as our first simulated example shows, it may suppress also the random spurious peaks caused by the red noise of stellar activity and thus enhance the detection of smaller planets.

We find that the partial periodograms are superior to classical periodograms, the PDC, and USuRPER, in their ability to separate shape and shift modulations. Moreover, the partial periodograms identify shape variability as the cause for periodic RV signals more reliably than analysis of BIS or FWHM data would. The main limitation we identified is that signals at identical periods cannot be perfectly separated if both shape and shift modulations are present in the data. Further analysis is required to fully characterize the behavior of the new partial periodograms, for example when multiple signals are present at close, but nonidentical, frequencies.

The emergence of the spurious peak in a period of $40$\,days in the S~Mus case raises the issue of the effects of sampling patterns. In other kinds of periodograms, especially Fourier-based periodograms such as the GLS, it is well known that sampling patterns (``window function'') might induce spurious peaks, for example, through aliasing. We plan to investigate how the distance-based periodograms (PDC, USuRPER, and the partial extensions presented here) are affected by those phenomena in the future.

The partial shift and shape periodograms offer a new approach to studying astronomical spectra, enabling researchers to distinguish stellar activity and pulsation from orbital RV shifts when characterizing observed systems\footnote{PDC, USuRPER and the partial distance correlation periodograms are available as part of the SPARTA package, at \url{https://github.com/SPARTA-dev/SPARTA}.}.

\begin{acknowledgements}
We are grateful to the anonymous referee for comments that helped improve the manuscript.
We acknowledge observational contributions by Nami Mowlavi, Lovro Palaversa, Kateryna Kravchenko, Pierre Dubath, Maroussia Roelens, and Berry Holl. The Swiss 1.2\,m Euler telescope is supported by the Swiss National Science Foundation. We thank Itamar Reis, Stav Bar-Sheshet and Dolev Bashi for reviewing and commenting on an early version of the manuscript. 
This work was supported by a grant from the Tel Aviv University Center for AI and Data Science (TAD).
RIA acknowledges support from the European Research Council (ERC) under the European Union's Horizon 2020 research and innovation programme (Grant Agreement No. 947660). RIA further acknowledges support through a Swiss National Science Foundation Eccellenza Professorial Fellowship (award PCEFP2\_194638). The research of SS is supported by Benoziyo prize postdoctoral fellowship.

The analyses done for this paper made use of the code packages: \texttt{Astropy} \citep{astropy1,astropy2}, \texttt{NumPy} \citep{numpy}, \texttt{PyAstronomy} \citep{pyast}, \texttt{SciPy} \citep{2020SciPy} and \texttt{SPARTA} \citep{SPARTA2020}.
\end{acknowledgements}

\bibliographystyle{aa}
\bibliography{pPDC}

\end{document}